\documentclass[12pt]{iopart}
\usepackage{epsfig}  
\begin{document}

\title{Probing small-x gluons
by low-mass Drell-Yan pairs 
at RHIC and LHC}

\author{George Fai\dag, Jianwei Qiu\ddag, and Xiaofei Zhang\dag\  
}

\address{\dag\  Center for Nuclear Research, Department of Physics,
Kent State University, 
Kent, Ohio 44242, USA}

\address{\ddag\ Department of Physics and Astronomy, Iowa State University,
Ames, Iowa 50011, USA}

\begin{abstract}
We calculate the transverse momentum distribution of low-mass Drell-Yan pairs 
in QCD perturbation theory with all-order resummation. 
We demonstrate that 
the transverse momentum distribution of low-mass Drell-Yan pairs is an 
advantageous source of constraints on the gluon distribution and its nuclear 
dependence.  With the reduction in background, we argue that low-mass 
Drell-Yan pairs in the forward region provide a good and clean probe 
of small-x gluons at RHIC and LHC.
      
\end{abstract}



\maketitle

Precise knowledge of the gluon distribution at small $x$ is critical 
for reliable predictions of important pp, pA and AA processes studied at
the Relativistic Heavy Ion Collider (RHIC)
and at CERN's Large Hadron Collider (LHC). 
Based on the data available, many models have been proposed to describe 
the nuclear parton distribution functions.  Different models
give very different predictions for the nuclear gluon distribution 
functions at RHIC and LHC energies\cite{Accardi:2003be}. The knowledge of
the nuclear gluon distribution is still very poor, and we need good 
processes to extract 
the gluon distribution function.
A good process to probe the gluon distribution must satisfy the following
criteria: (i) it has to be reliably calculable via 
perturbative quantum chromodynamics (pQCD), which requires the 
process to be pQCD factorizable;  
(ii) the production cross section must be dominated by gluon-initiated 
sub-processes;  (iii) final state 
medium effects must be small; and also (iv) the process should 
have sufficient production rate to be observed.
We show in this contribution that  
low-mass lepton pairs produced at large transverse momentum
and rapidity provide such a good probe for
the nuclear gluon distribution function\cite{Berger:1998ev}. 
Drell-Yan production is also free from the 
complications of photon isolation that beset 
studies of prompt photon 
production\cite{Berger:1998ev,Berger:1990es}.

In the Drell-Yan process, the massive 
lepton pair is produced via the 
decay of an intermediate virtual photon, $\gamma^*$.  Within the context of 
pQCD,
the Drell-Yan cross section in 
a collision between hadrons $A$ and
$B$, $A(P_A)+B(P_B)\rightarrow \gamma^*(\rightarrow l\bar{l}(Q))+X$,
can be expressed in terms of the cross section for production of an 
unpolarized virtual photon of the same invariant mass\cite{Berger:1998ev}, 
\begin{equation}
\frac{d\sigma_{AB\rightarrow \ell^+\ell^-(Q) X}}{dQ^2\,dQ_T^2\,dy}
= \left(\frac{\alpha_{em}}{3\pi Q^2}\right)
  \frac{d\sigma_{AB\rightarrow \gamma^*(Q) X}}{dQ_T^2\,dy}\, \,\, .
\label{DY-Vph}
\end{equation}
The variables $Q$, $Q_T$, and $y$ are the invariant mass, transverse momentum, 
and rapidity of the pair. Symbol $X$ stands for an inclusive sum over final 
states that recoil against the virtual photon. 
An integration has been performed over the angular 
distribution in the lepton-pair rest frame. 

Recently, it was shown that the Drell-Yan cross section at $Q\ll Q_T$ can be 
factorized as\cite{Berger:2001wr}
\begin{eqnarray}
\frac{d\sigma_{AB\rightarrow \gamma^*(Q) X}}{dQ_T^2\,dy}
&=&\sum_{a,b} \int dx_1 \phi_{a/A}(x_1,\mu) 
              \int dx_2 \phi_{b/B}(x_2,\mu)
\nonumber\\&\ &
 \times \Bigg\{ 
 \frac{d\hat{\sigma}_{ab\rightarrow \gamma^*(Q) X}^{(Dir)}}{dQ_T^2\,dy}
        (x_1,x_2,Q,Q_T,y;\mu_F,\mu)
\label{DY-mfac} \\
&\ & {\hskip 0.1in} +
\sum_c \int\frac{dz}{z^2} \left[
 \frac{d\hat{\sigma}_{ab\rightarrow c X}^{(F)}}{dp_{c_T}^2\,dy}
        (x_1,x_2,p_c=\frac{\hat{Q}}{z};\mu_F,\mu) \right]\nonumber\\
&\ & {\hskip 0.2in}D_{c\rightarrow\gamma^*(Q) X}(z,\mu_F^2;Q^2)
 \Bigg\} \, .
\nonumber 
\end{eqnarray}
In the above equation,  
$d\hat{\sigma}_{ab\rightarrow c X}^{(F)}/dp_{c_T}^2\,dy$ 
and $d\hat{\sigma}_{ab\rightarrow \gamma^*(Q) X}^{(Dir)}/dQ_T^2\,dy$, 
are perturbatively calculated short-distance hard parts. The sum 
$\Sigma_{a,b}$ runs over all parton flavors;  $\phi_{a/A}$ and
$\phi_{b/B}$ are parton distribution functions; $\mu$ is the renormalization 
and the factorization scale, with $\mu_F$ representing the fragmentation 
scale.
The superscripts (Dir) and (F) indicate direct contribution and fragmentation
contribution, respectively. The $D_{c\rightarrow\gamma^*(Q) X}$ are 
resummed  virtual photon fragmentation functions, including
all order large $\ln^m(Q_T^2/Q^2)$ contributions in the region $Q_T^2\gg Q^2$. 
The four-vector $\hat{Q}^\mu$ 
is defined to be $Q^\mu$ but with $Q^2$ set to be zero. 

\begin{figure}
\begin{minipage}[c]{7.6cm}
\centerline{\includegraphics[height=7cm,width=8cm]{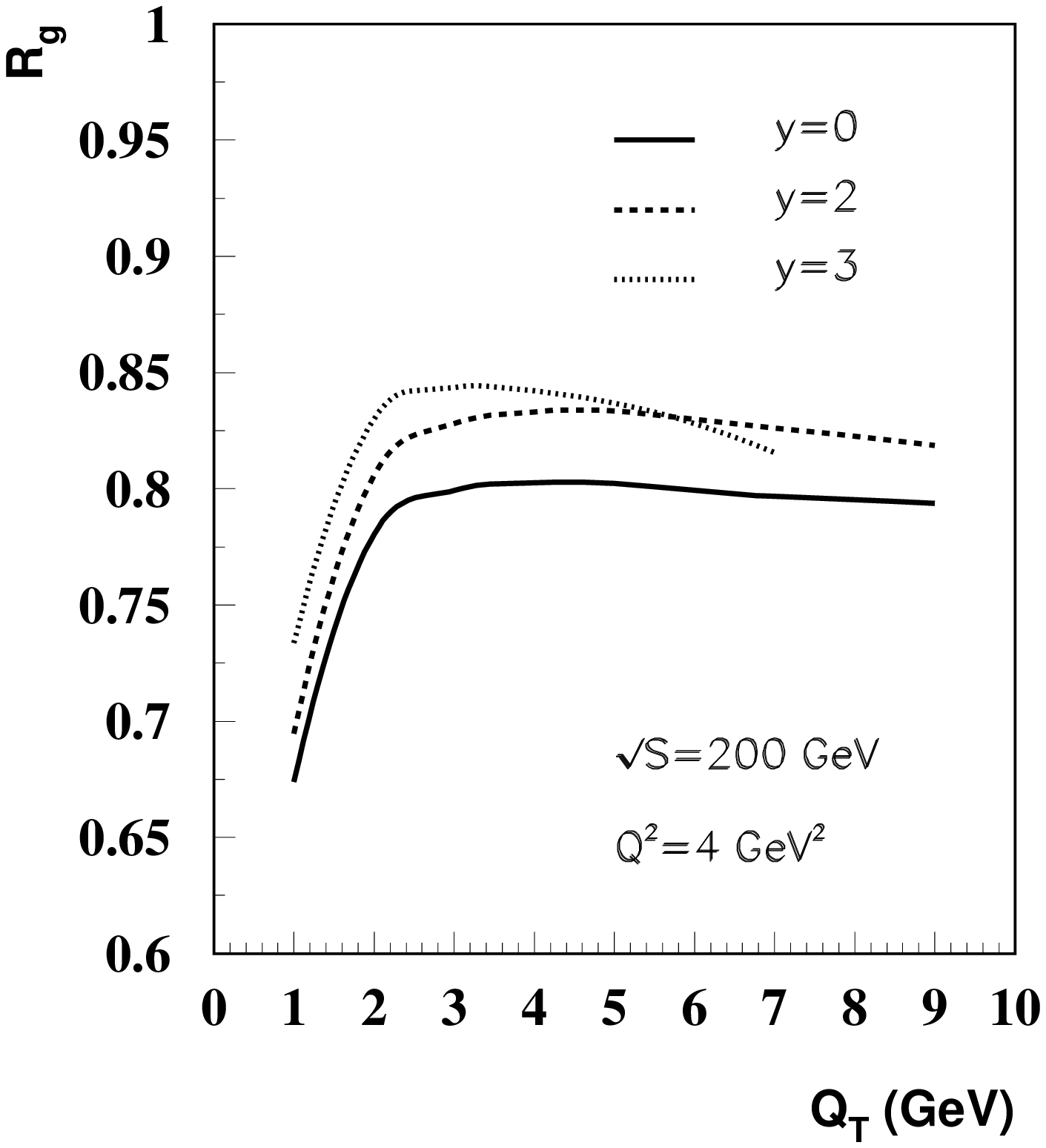}} 
\label{fig1(a)}
\end{minipage}
\hfill
\begin{minipage}[c]{7.6cm}
\includegraphics[width=8cm,height=7cm]{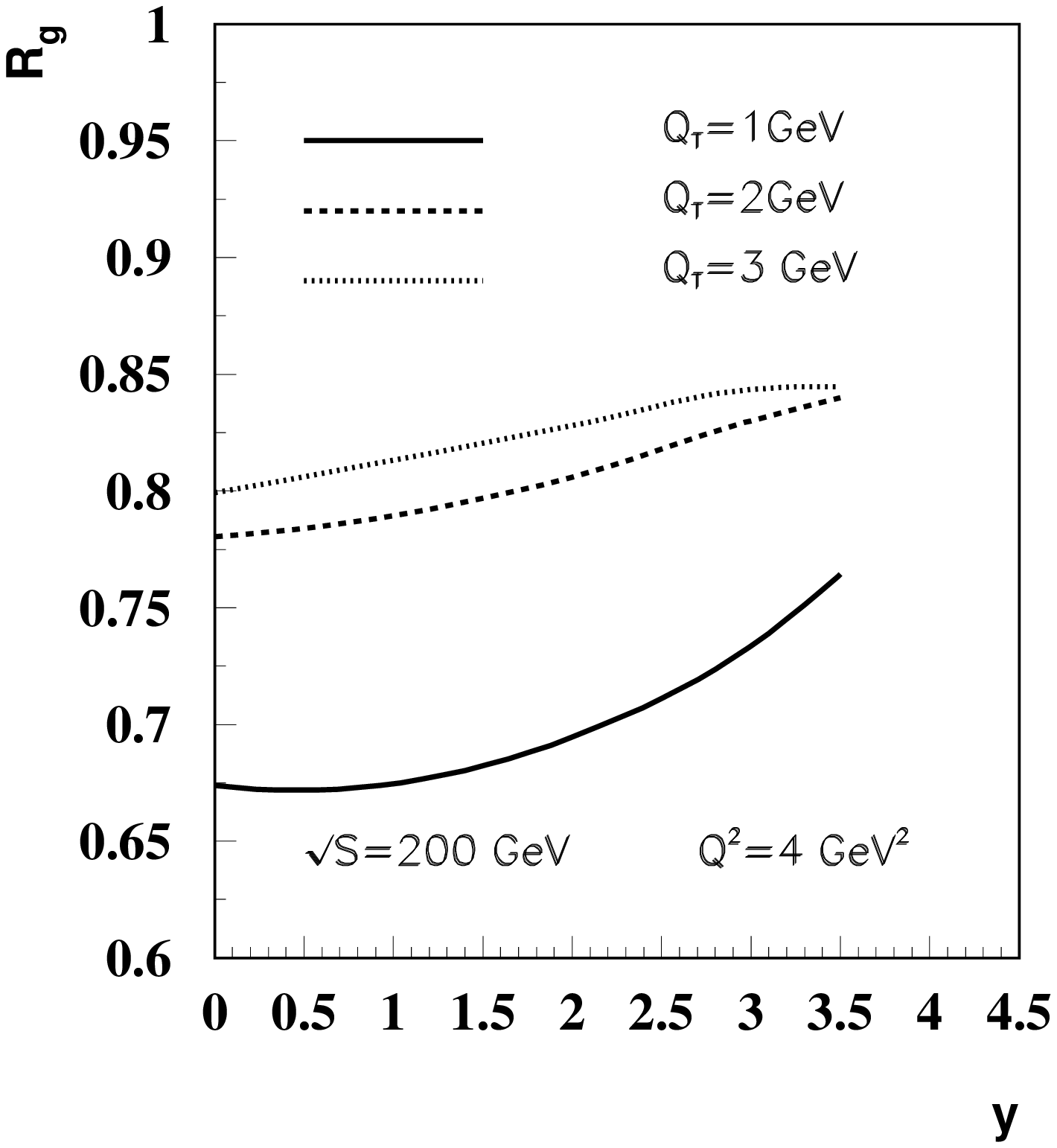} 
\vspace{-0.3in}
\label{fig1(b)}
\end{minipage}
\vspace{-0.1in}
\caption{The ratio $R_{g}$ defined in Eq.~(\protect\ref{R-g})
with $Q^2=$4 GeV$^2$. Left panel: $R_{g}$ as function of $Q_T$ for different
rapidities; right panel: $R_{g}$ as a 
function of rapidity y for different transverse momenta.}
\end{figure}

To demonstrate the relative size of gluon initiated contributions, we 
define the ratio
\begin{equation}
R_g = \left.
\frac{d{\sigma}_{AB\rightarrow \gamma^*(Q) X}(\mbox{gluon-initiated})}
     {dQ_T^2\,dy} \right/
\frac{d{\sigma}_{AB\rightarrow \gamma^*(Q) X}}{dQ_T^2\,dy}\, \,\, .
\label{R-g}
\end{equation}
The numerator includes the contributions from all partonic sub-processes with 
at least one initial-state gluon, and the denominator includes 
all sub-processes. 
In Fig.~1, we show $R_g$ as a function of $Q_T$ at fixed $y$ (left panel),
and as a function of $y$ at fixed $Q_T$ (right panel)
at $Q=2$~GeV for RHIC energy. CTEQ5M parton distribution functions are used 
in all our calculations reported here,
and we set all scales to $\sqrt{(Q_T^2+Q^2)}$. 
Fig.~1 confirms that gluon-initiated sub-processes dominate the Drell-Yan cross section and thus 
low-mass Drell-Yan lepton-pair production at large transverse momentum 
is an excellent source of information on the gluon distribution\cite{Berger:1998ev}.
The fall-off of $R_g$ at very large $Q_T$ is related to the reduction of phase 
space and to the fact that the cross sections are evaluated at larger 
values of the partons' momentum fractions.

However, as seen from Eq. (1),
there is a phase space penalty associated with the finite mass of 
the virtual photon, and the Drell-Yan factor $\alpha_{em}/(3\pi
Q^2)< 10^{-3}/Q^2$ renders the production rates for massive
lepton pairs small at large values of $Q$ and $Q_T$.  In order to enhance the
Drell-Yan cross section while keeping the dominance of the gluon
initiated sub-processes, it is useful to study lepton pairs with low
invariant mass and relatively large transverse momenta\cite{Berger:2001wr}.  
With the large transverse momentum $Q_T$ setting the hard scale of the 
collision, the invariant mass of the virtual photon $Q$ can be small,
as long as the process can be identified experimentally, and the numerical 
value $Q\gg\Lambda_{\rm QCD}$.  For example, the cross section for Drell-Yan
production was measured by the CERN UA1 Collaboration~\cite{UA1-Vph} 
for virtual photon mass $Q\in [2m_\mu, 2.5]$~GeV.  

\begin{figure}
\begin{minipage}[c]{7.6cm}
\centerline{\includegraphics[width=8cm,height=8cm]{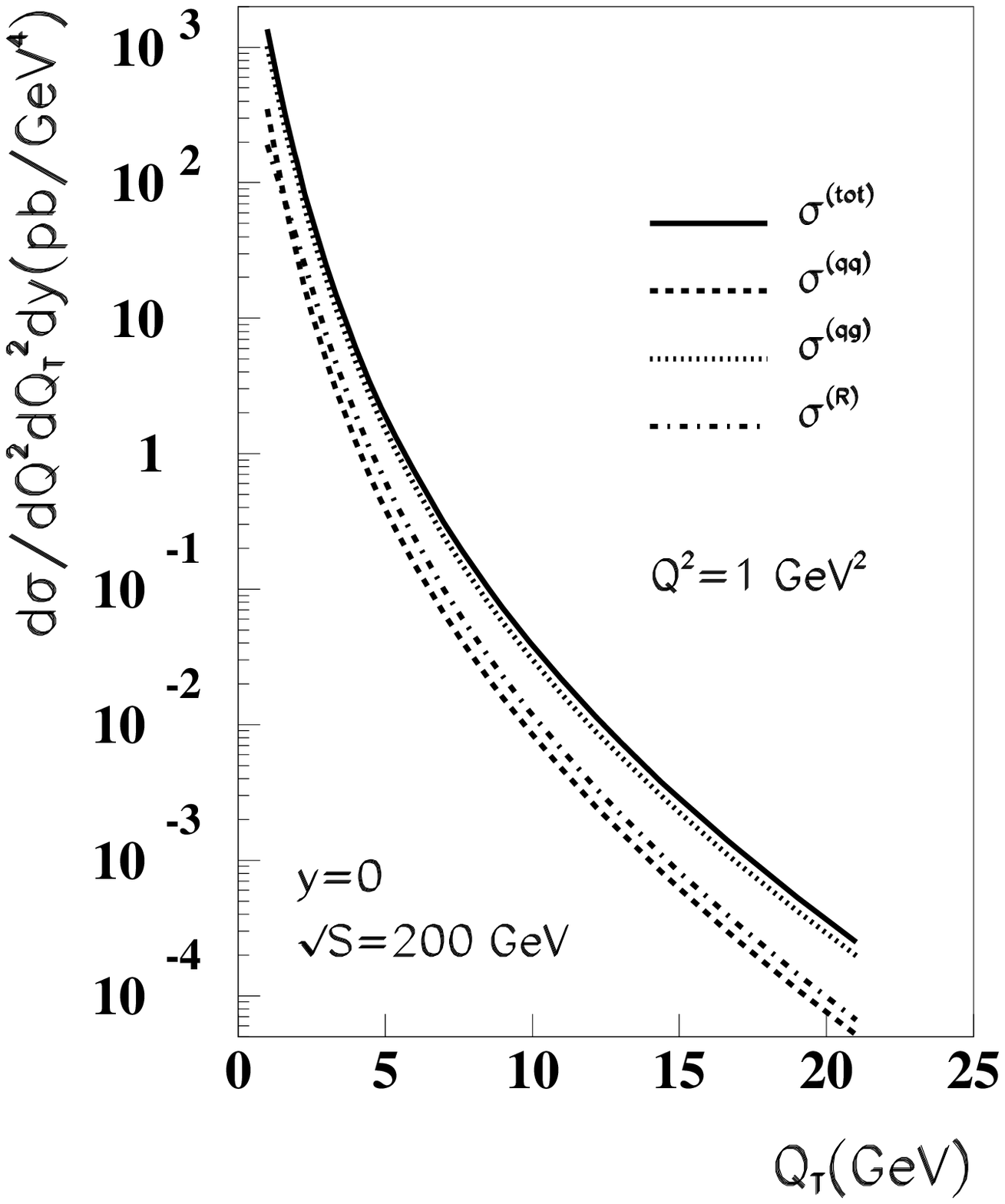}} 
\vspace{-0.1in}
\label{fig2(a)}
\end{minipage}
\hfill
\begin{minipage}[c]{7.6cm}
\vspace{0.0in}
\includegraphics[width=8cm,height=8cm]{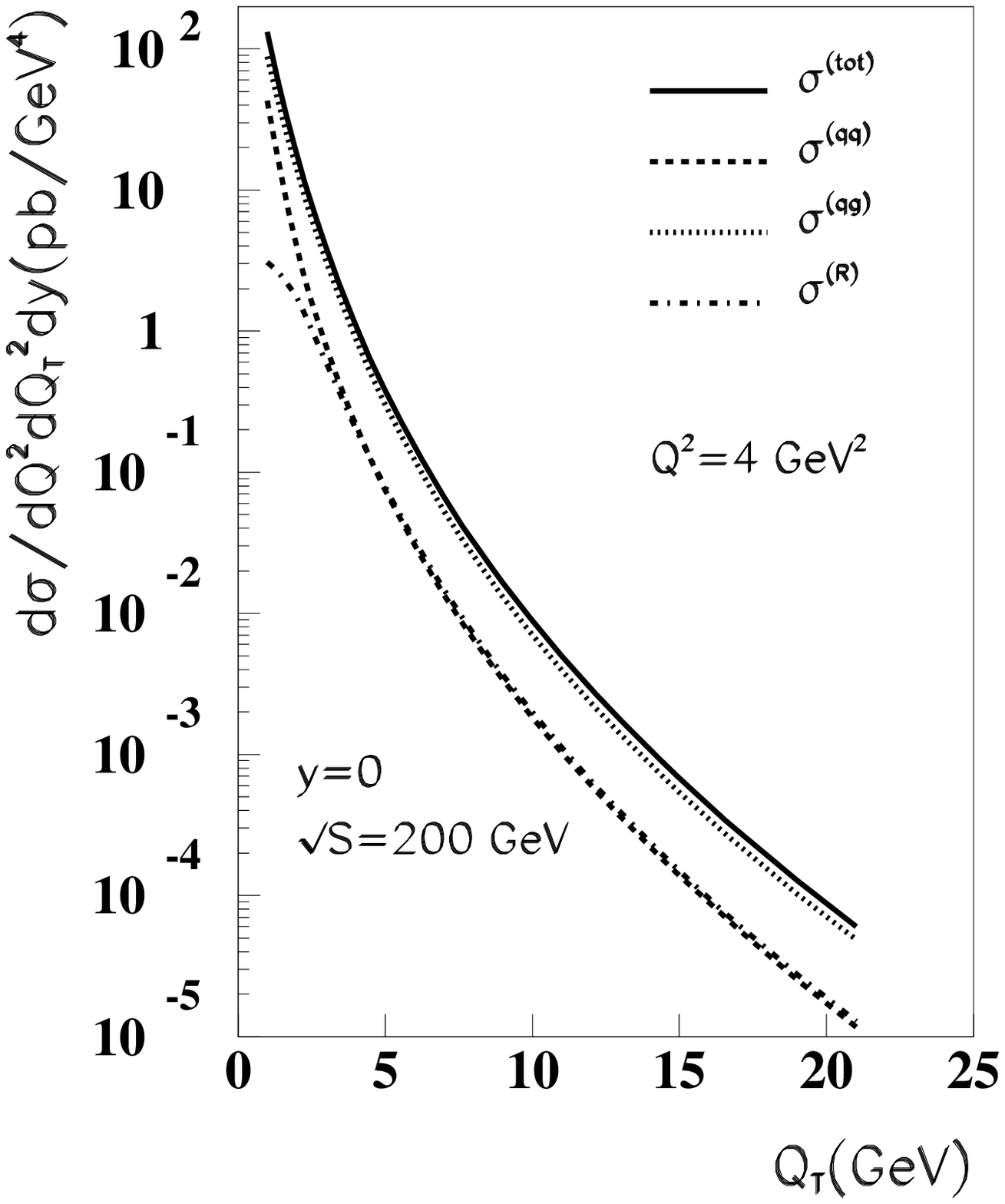} 
\vspace{-0.3in}
\end{minipage}
\vspace{-0.1in}
\caption{Drell-Yan cross section as a function of $Q_T$ at RHIC
energy $\sqrt{s}=200$ GeV and rapidity $y=0$, for mass 
$Q=1$ GeV (left panel) and $Q=2$ GeV (right panel). The top lines
are the total cross sections; different contributions are also shown.
}
\end{figure}

Fig.~2 shows the all-order resummed result
for the Drell-Yan cross section as a function of $Q_T$ at the RHIC
energy of $\sqrt{s}=200$ GeV and rapidity $y=0$ for two values of the mass, 
$Q=1$ GeV (left panel) and $Q=2$ GeV (right panel). The cross section 
increases by about a factor
10 when the Drell-Yan mass decreases from 2 GeV to 1 GeV.
It might still be a challenge to 
measure the low-mass Drell-Yan production with the production rate shown
in Fig.~2. At the LHC, both the collision energy and luminosity are 
significantly improved. Our calculation shows that  
the production rate at the LHC is sufficient. 

It is very important to estimate the region of $x$ in the gluon 
distribution probed by the low-mass Drell-Yan process.
The $x$ integration of the Dell-Yan cross section runs from $x_{min}$
(given by the kinematics)
to 1. We introduce a cutoff $x_{cut}$ to limit the integration to range 
$x_{min}$  to $x_{cut}$ for the parton in the ``target'' beam, 
and define the ratio
\begin{equation}
R_x = \left.
          \int_{x_{\rm min}}^{x_{\rm cut}} dx_2
          \left( \frac{d\sigma^{\rm DY}}{dx_2} \right)
          \right/
          \int_{x_{\rm min}}^{1} dx_2
          \left( \frac{d\sigma^{\rm DY}}{dx_2} \right)  \,\,\, .
\label{Rx}
\end{equation}
The shape of $R_x$ establishes which region dominates the $x$ 
integration ---  the region of $x$ low-mass Drell-Yan data 
could provide precise information about.
In Fig.~3 we plot $R_x$ as a function of $x_{cut}$ for $Q=1$ GeV 
(left panel) and $Q=2$ GeV (right panel)
for different rapidities and transverse momenta 
$Q_T$.
Fig.~3 shows that
in the central rapidity 
region $x_2\sim [10^{-2}, 10^{-1}]$
dominates the integration  for
both $Q=1$ GeV and $Q=2$ GeV. 
Fig.~3 also provides important information about the forward region. 
At $y=2$, 90\% of the cross-section is given by $x_2 < 0.01$,
which is exactly the shadowing region of the
nuclear parton distribution function. This region dominates the integration  
for both $Q=1$ GeV and $Q=2$ GeV for $y=2$. 
Our calculation for the rapidity distribution 
of the total cross section shows that the cross section does not start to 
drop until $y\sim 3$. 
This indicates 
that the production rate in the forward region
is not smaller than in the central region as long as $y\le 3$. 
Therefore, low-mass Drell-Yan in the forward region is an
excellent probe of the nuclear shadowing effect and may help us understand 
the suppression of hadron production in the forward region at RHIC.

\begin{figure}
\begin{minipage}[c]{7.6cm}
\centerline{\includegraphics[width=8cm,height=7cm]{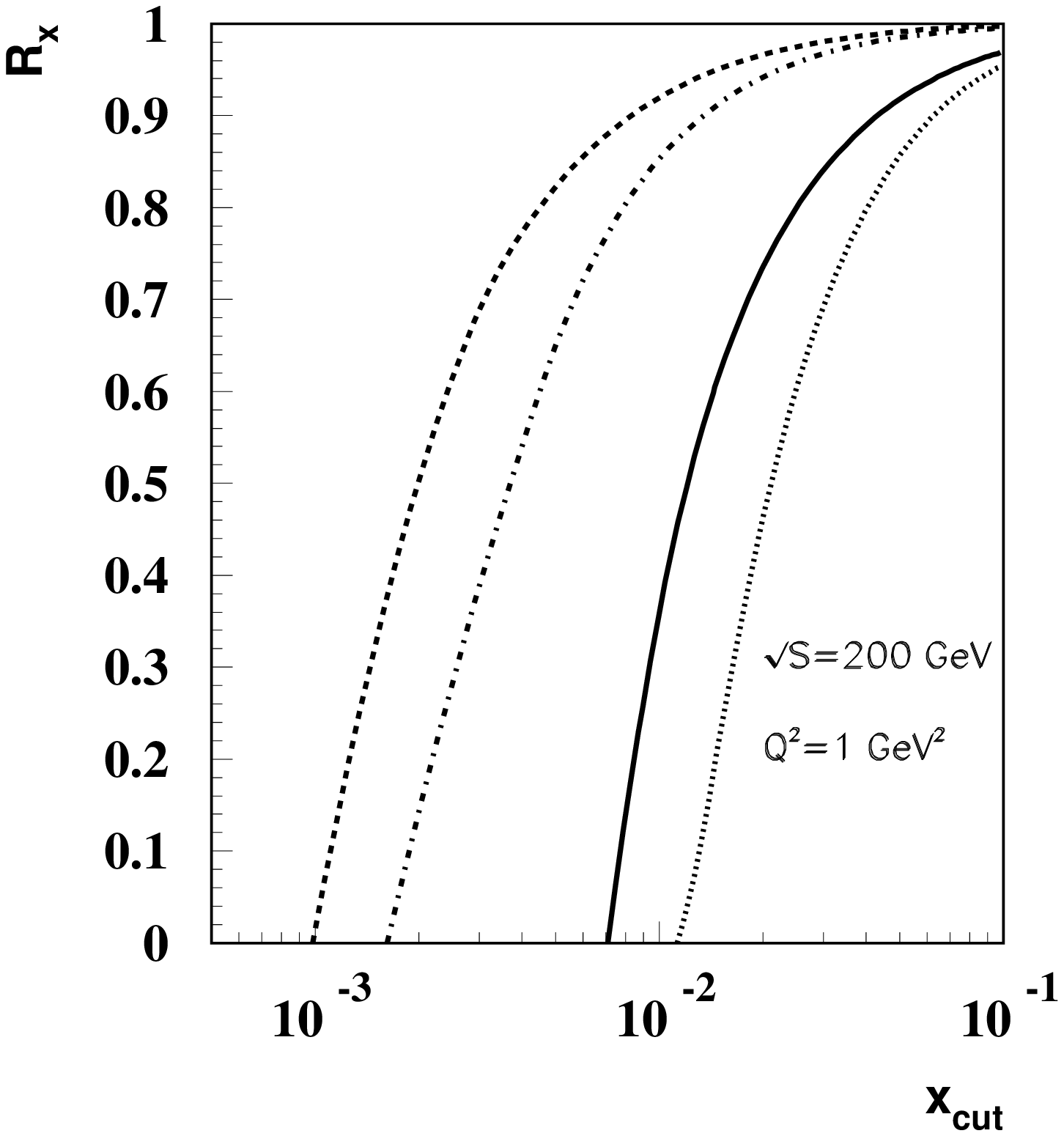}} 
\label{fig3(a)}
\end{minipage}
\hfill
\begin{minipage}[c]{7.6cm}
\includegraphics[width=8cm,height=7cm]{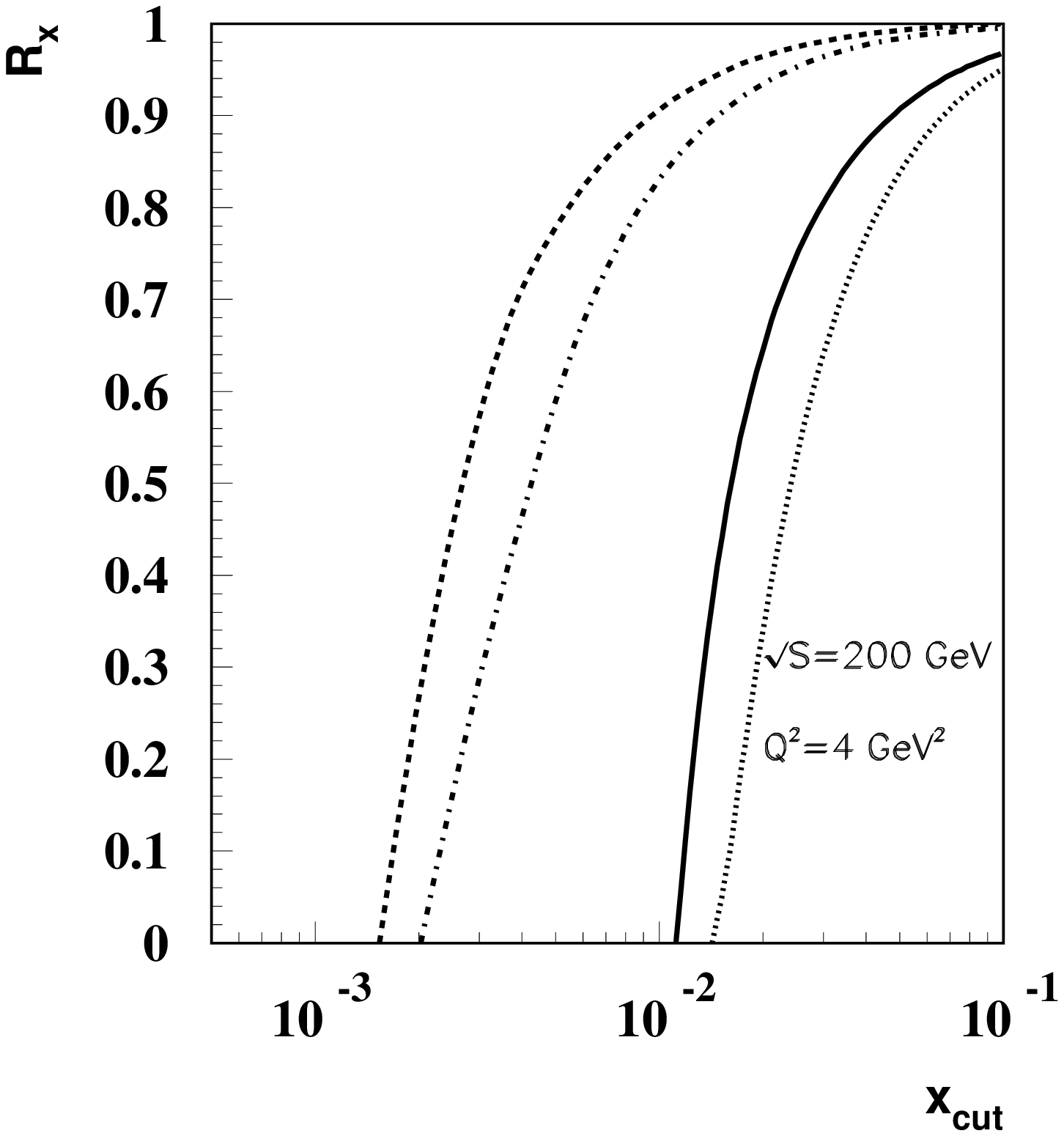} 
\end{minipage}

\caption{The ratio $R_{x}$ defined in 
Eq.~(\protect\ref{Rx})
for $Q=1$ GeV (left panel) and $Q=2$ GeV (right panel). 
Different lines in the figure are the  $R_{x}$ at different $y$
and $Q_T$: $y=0, Q_T=1$ GeV (solid); $y=0, Q_T=2$ GeV (dotted);
$y=2, Q_T=1$ GeV (dashed); $y=2, Q_T=2$ GeV (dot-dashed).}
\end{figure}

In summary, low-mass Drell-Yan dilepton production 
is a potentially clean probe of small-$x$ gluons, without strong
final-state interaction.  When $Q_T>1$ GeV,   
the gluon initiated sub-processes contribute more than 
70\% to the cross section.
We have also shown that the forward region is very sensitive to 
small-$x$ gluons. Unfortunately, low-mass Drell-Yan dilepton production 
suffers from low production rate at RHIC 
due to the Drell-Yan factor. 
At  LHC energies, high-$Q_T$ low-mass Drell-Yan production 
is an excellent probe of 
gluon distributions.

\vskip .5 cm

This work was supported in part by the U.S. Department of Energy,
under grants DE-FG02-86ER40251 and DE-FG02-87ER40371.


\vspace{0.5cm}
\begin{thebibliography}{4}
\bibitem{Accardi:2003be}
A.~Accardi {\it et al.},
arXiv:hep-ph/0308248.

\bibitem{Berger:1998ev}
E.~L.~Berger, L.~E.~Gordon and M.~Klasen,
Phys.\ Rev.\ D {\bf 58}, 074012 (1998)
[arXiv:hep-ph/9803387].

\bibitem{Berger:1990es}
E.~L.~Berger and J.~w.~Qiu,
Phys.\ Lett.\ B {\bf 248}, 371 (1990).

\bibitem{Berger:2001wr}
E.~L.~Berger, J.~w.~Qiu and X.~f.~Zhang,
Phys.\ Rev.\ D {\bf 65}, 034006 (2002)
[arXiv:hep-ph/0107309].
\bibitem{UA1-Vph}
C. Albajar {\it et al.}, UA1 Collaboration, Phys. Lett. {\bf B209},
397 (1988).

\end{thebibliography}
\end{document}